\documentclass[12pt]{article}
\usepackage{fullpage, natbib, url, authblk}
\usepackage{amsmath, amssymb, amsfonts}
\usepackage{rotating, caption, subcaption, wrapfig, multirow, floatrow}
\usepackage{graphicx}
\usepackage[svgnames]{xcolor}

\title{Converting College Football Point Spread Differentials to Probability}
\author[1]{Ryan Sides*}
\author[2]{Jane L.~Harvill}
\author[3]{Victoria R.~Sides}
\affil[1]{\protect\raggedright
Texas Woman's University, Division of Mathematics, Denton, TX, U.S., e-mail: rsides@twu.edu}
\affil[2]{\protect\raggedright 
Baylor University, Department of Statistical Science, Waco, Texas, U.S., e-mail: jane\_harvill@baylor.edu}
\affil[3]{\protect\raggedright 
Baylor University, Department of Statistical Science, Waco, Texas, U.S., e-mail: victoria\_sides1@baylor.edu}
\date{\today}
\begin{document}

\maketitle

\section{Introduction}
\label{sec:introduction}
Analytics have become mainstream in almost every industry, and sports are no exception. In particular, sports betting
has historically relied heavily on analytics. 
A conservative estimate of money bet on college football alone is in excess of one billion dollars annually~\citep{purdam2020}. And so while much work has been done in this area, most of it is proprietary due to the competitive
profit-driven nature of this big business.  
The primary objective of this paper is to provide a method for sports bettors to determine if they have a positive expected value bet based on the betting lines available to them and how they think the game will play out.  Methods and a publicly available online tool are provided to answer questions like, ``If a model, or aggregate of models, says Team A wins by 7.9 points, should I bet that team to win by seven or more?'' 

The outline of the remainder of this paper is as follows.  A basic overview of sports betting is provided in Section~\ref{sec:basics}.  Following that,   Section~\ref{sec:approaches} describes one approach to creating a projected spread before discussing multiple techniques and their shortcomings for translating this to a betting edge. In Section~\ref{sec:new}, a method is presented that combines modifications of those techniques to arrive at an improved solution for determining the probability a bettor will win a specific bet based on his belief about that game.  In Section~\ref{sec:future}, some potential extensions are briefly discussed.

\section{The Basics of Sports Betting}
\label{sec:basics}
A successful sports gambler will consistently make positive expected value plays, making picks that will yield a profit over time.  Betting on a team to win a game, known as betting a moneyline, is popular in baseball and hockey, though less so in basketball and American football, referred to here as ``football.''  When betting moneyline markets, to achieve positive expected value, a bettor needs to have an objective, consistent, and repeatable approach to assessing the probability that each team wins.  When betting moneylines, a relatively simple math problem can be solved to determine a threshold for which advantages are worthy of an investment. So the bettor needs only to decide their personal threshold and how much money they are willing to bet.

For example, consider a $-120$ bet using American odds.  In this schema, an individual bets \$120 dollars on a team to win.  If the team wins, the bettor receives back the \$120 and an additional \$100.  American odds can be converted to what is a ``break-even percentage'' $p$ using 
\begin{equation}
    \label{eq:odds}
    p = 100\left(\frac{|\min(100,\textrm{odds})|}{100 + |\textrm{odds}|}\right)\%.
\end{equation}
Another percentage is the ``cover percentage,'' which is the percentage of times a bet is won. This is easy to describe and possible to quantify after the game is played.  However, a priori, finding the cover percentage is one of the most challenging problems in sports betting. The ``betting edge,'' or simply ``edge,'' is defined as the difference between the cover percentage and the break-even percentage.  Negative edge values imply that the model or belief system used indicates no wager should be made. 
All bettors bet under the paradigm that the positive edge must be large enough to justify the bet over the risk.  However, the threshold for how large the positive edge should be to place a bet is subjective.

If the bettor bets the moneyline in this same scenario over a long period of time, to achieve a positive return using equation~\eqref{eq:odds}, that side must have at least a 54.5\% chance of winning over that period of time.  For a single game, if the bettor believes the team has a greater than 54.5\% chance of winning, betting the moneyline is a wise bet.  The bet is a positive expected value play under the bettor's belief system.  

The sportsbooks, which are establishments that accept bets, charge a commission referred to as the ``juice'' or ``vig,'' short for vigorish.  The juice varies based on the odds.  Aside from promotions, the juice is always there, even when it's not apparent.  Suppose in the previous example that the opposing team's moneyline is $+110$, meaning that a \$100 bet would win \$110.  Using equation~\eqref{eq:odds}, the bet needs to hit 52.4\% of the time for a bettor to be profitable.  This creates a situation where some games provide little or no value to the bettor.  If the belief is that the favored team will win 54\% of the time, neither side is profitable in a long-run situation.  This is because the 54\% win percentage falls below the break-even threshold of 54.5\% for the favored team and short of the 47.6\% break-even threshold for the underdog.  Further, if the bettor believes the team will win 55\% of the time, it is arguable that the expected return on a winning bet is so small, \$5.50 from \$120 investment, that it is not worth the capital invested.

Achieving positive expected value play becomes much more complicated when a bet is based on a point spread. A point spread bettor concerns himself with the margin of victory or defeat.  To win a bet in the point spread market, the team that is bet on must win by at least a specified number of points, or lose by less than that.  This is the most common type of bet in football.  Converting a point spread into a simple win probability is a relatively simple task (see, for example, {\tt boydsbets.com/nfl-spread-to-moneyline-conversion/}).  However, translating the difference between a point spread and a bettor's projected point spread is a much more complicated problem. As previously mentioned, most of the work that has been done to assess the value of these differences is proprietary.

Further complicating this problem is that football games tend to end with common point differentials, like three points or seven points.  On the other hand, it is unusual for a football game to end with a point differential of five points.  Consequently, a bettor who believes a team will win by six points, but can bet a win by four or more, has a smaller advantage than one who believes a team will win by four points, but can bet them to win by two or more.  Thus, how does a sports bettor take a specific point spread situation and determine whether betting is in their favor?  In other words, ``For the game of interest, does the bettor have a positive expected value play based on the available betting lines and the bettor's belief about how the game will play out?''

One technique that attempts to answer this is the Pythagorean expectation.  Originally developed by Bill James for season-long win percentages in baseball, the technique has been modified and applied to other sports.  For football, the modified version is often referred to as the Pythagorean wins and is defined as a fraction multiplied by $N =$ number of games played.  For football, 
\begin{equation*}
    {\mbox{Pythagorean wins}} = N\left(
    \frac{{\mbox{points for}}^r}{{\mbox{points for}}^r + {\mbox{points against}}^r}\right).
\end{equation*}
For the National Football League (NFL), for $N = 17$ games played, the suggested value of the exponent $r$ is 2.37.  One of the problems with using  Pythagorean wins is the subjective determination of the exponent.  Another problem is that the method depends on points scored by both teams and is a long-run expectation.  Consequently, it is better suited for a full season rather than a single game, as discussed by \citet{FootballOutsiders}, among others. 

The standard approach for determining if a single game bet is advantageous when betting with a point spread was developed by \citet{Stern}.  \citeauthor{Stern} uses a normal distribution with a standard deviation of 13.86\footnote{More recent values of the standard deviation are 13.5 points.} points for NFL games to find the probability that a team favored by a specific amount will win by a certain margin.  The potential flaw with this approach is that point differentials for football are discrete, and  the normal distribution determines probabilities of continuous-valued variables.  A more serious problem with \citeauthor{Stern}'s approach is the assumption that a team favored by five will win by exactly five at the same rate that a team favored by seven wins by exactly seven.  This assumption is a direct contradiction to reality; a score differential of five points is much less common than a score differential of seven.  While \citeauthor{Stern}'s approach is the basis for many pregame win probability models, including those at \citet{ProFootballReference}, it can be improved upon by incorporating the higher likelihood of certain score differentials in football games. And while there are many efforts to assess the win probability given a specific point spread, see for example \citet{Huggins}, efforts to quantify the betting edge between a projected point spread and one available to the gambler have not advanced to the public beyond the use of a normal distribution.

In what follows, a method is provided for sports bettors to determine if they have a positive expected value bet based on the betting lines available to them and how they think the game will play out.  Methods and a publicly available online tool are provided to answer questions like, ``If a model, or an aggregate of models, says Team A wins by 7.9 points, should I bet that team to win by seven or more?'' 

\section{Investigating Simple Approaches for College Football Betting Edges}
\label{sec:approaches}
One of the keys to profitability when betting a point spread over the long term is having a reliable projection system.  The basic idea is for the gambler to assess what he thinks the point spread should be for a given game. A bettor's point spread for a given game can be based on a power rating system, his own qualitative research, or some combination of these.  There are many popular publicly available power ratings. The work in this paper adopts Bill Connelly's SP+, currently available via ESPN+, to illustrate the approaches presented.  Connelly's system follows a type of Bayesian paradigm, beginning with some set of data based on at least one previous season's  games, updating throughout the current season to produce a power rating for every team.  Home-field advantage may be accounted for, which is usually agreed to be somewhere around 2 to 2.5 points in favor of the home team.  After accounting for home-field advantage, the difference between the two teams' ratings results in a projected point spread.  Hereafter, this is referred to as the ``system projected point spread,'' or ``system point spread.'' Serious bettors consider the information in a projection system like SP+ when deciding when to place a wager.  However, they also take into account their own research and knowledge, which may include things like injuries, transfers, or suspensions.  Most projection systems may not fully capture such team-specific real-time information.  Combining all this, bettors determine their projected point spread, referred to as the ``bettor's projected point spread'' or ``bettor's point spread.''

Every college football game has a ``betting point spread'' as determined by the bookmakers.  This betting point spread is often seen in any mention in the media for a game.  Attached to that betting point spread are odds, as discussed in Section~\ref{sec:basics}.  One common approach is to bet on any team that shows at least a two-point differential from the bettor's projected point spread and the betting point spread.  This approach is simple.  However, this to has the potential to be misleading since all final score differences in football aren't the same, which was also discussed in Section~\ref{sec:basics}.  Because points happen in three and usually seven point chunks, teams make decisions in the second half of play in order to get ahead of, or behind by, certain key numbers. Thus, there is a need to properly assess whether the difference between the bettor's point spread and betting point spread matters, as well as if the edge is large enough to be worthy of an investment.  

\subsection{Using a Normal Distribution to Quantify Betting Edge}
\label{sec:normal}
\begin{figure}[!ht]
  \caption{Bubble plot of point spread vs.~score differential for 2021 college football games.}
  \includegraphics[width=0.9\textwidth]{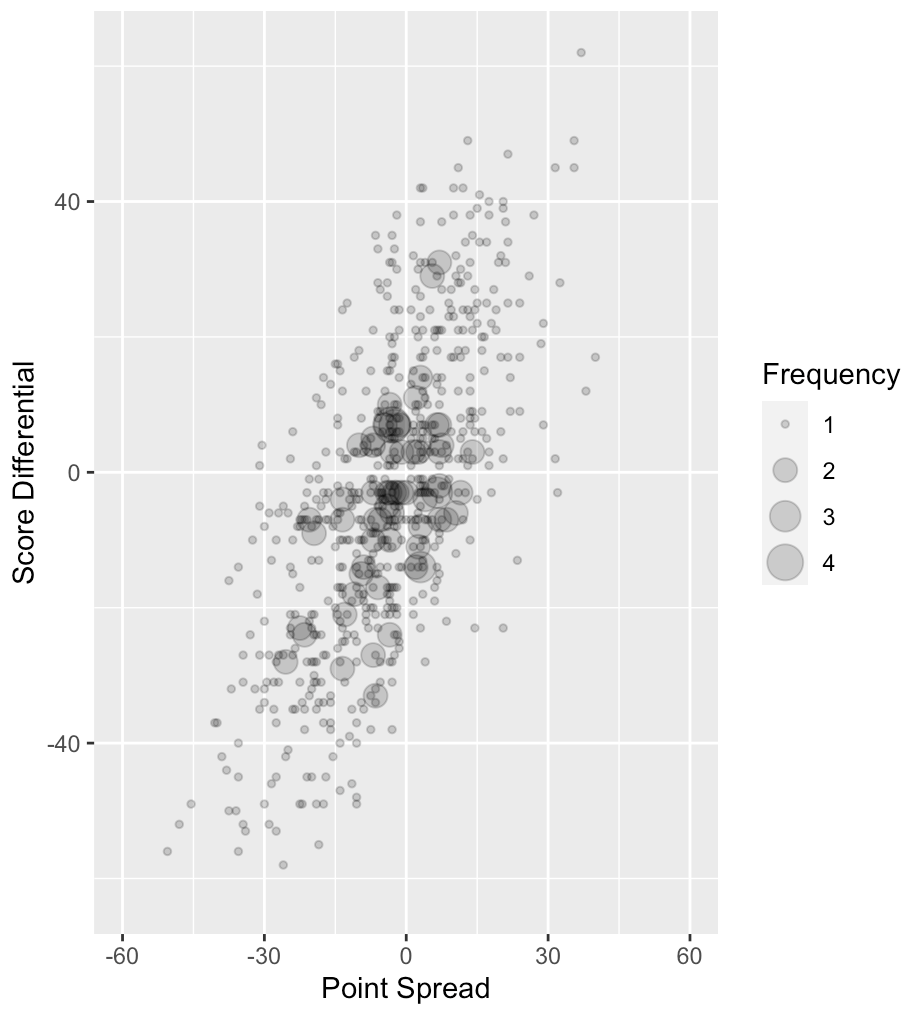}
  \label{fig:plot}
\end{figure}

For all college football games, the standard deviation of the score differential is slightly greater than 20 points. However, for games with similar point spreads, the conditional standard deviation decreases to around 15.  For the 2021 season, the standard deviation of the score differential for all games was 21.01; for games with similar point spreads, the standard deviation was 15.35.  One way to interpret the conditional standard deviation of 15 is as follows: for all teams projected to win by 6.5 points, only a few of them would win by more than $2 \times 15 + 6.5 = 36.5$ or lose by more than $2 \times 15 - 6.5 = 23.5$.  During the 2021 season, there were 82 games with a point spread of six to seven.  For these games, only three (3.7\%) had a win of more than 36 points or a loss of less than 23.  Figure~\ref{fig:plot} is a bubble plot of the score differential versus the point spread for all games of the 2021 season.  Each point represents the score differential for at least one game with a specific point spread.  Points with a larger physical size represent more than one game.  The plot's legend indicates how many games are represented by the various plotting symbol sizes.

A basic approach for finding the probability of covering the projected point spread for a specific game was discussed in Section~\ref{sec:basics}.  Because the projected point spreads from sportsbooks or models can be believed to be relatively accurate, an often-suggested starting point for finding the probability is to use a normal distribution with the mean being the projected point spread and a standard deviation of 15.\footnote{The Skellam distribution is sometimes used for finding these probabilities, as it deals with the difference in Poisson distributions. However, this distribution is only appropriate for sports like hockey and baseball where scoring happens one point at a time.}  To illustrate, consider the game on October 30, 2021, when the University of Texas traveled to Waco, Texas to play Baylor University.  The closing line had Baylor as a 2.5-point favorite.  The projected point spread coming from SP$+$ was Baylor $-2.9$. If $\Phi$ represents the normal probability, then
$$
P({\mbox{Baylor covers}}) = \Phi\left(\frac{-2.5-(-2.9)}{15}\right) = 0.5106;
$$
that is, using the normal distribution with a mean equal to the projected point spread of $-2.9$ and a standard deviation of 15, the probability of Baylor covering the 2.5 point spread is 0.5106.

Using this method, SP$+$ suggests that neither side is a profitable wager.  Assuming the standard $-110$ wager, neither team has a cover probability larger than 52.4\%.  However, because the most common football point differential is three and a tie is impossible, it is clear that outcomes around zero are too heavily weighted, and outcomes around three points are not weighted heavily enough.  Consider, for example, the following two scenarios.  The first is a bettor's believed spread of 2.5 versus a betting spread of 3.5.  The second is a bettor's believed spread of 4.5 versus a betting spread of 5.5.  Under the methodology just described, these two scenarios are treated as equivalent, even though they are not.  A college football game is much more likely to end with a point differential of three instead of five, making the former a larger betting edge.  This equal weighting of nonequivalent events is a flaw that will be addressed.

When the point spread is zero or is not an integer, there are two results that can occur; one team covers or the other team covers.  In the Texas vs.~Baylor example, the point spread was Baylor by 2.5.  The probability that Texas covers is one less than the probability Baylor covers.  On the other hand, if the point spread is an integer, there exists the possibility of a ``push''; that is, the score differential ends exactly on the spread.  In this case, three possible outcomes and a wide variety of methods and philosophies for dealing with a push.  These are not included in this paper.

\subsection{Using Historical Data to Quantify Betting Edge}
\label{sec:historical}
Historical data can be used to weight each point differential to overcome equal weighting of nonequivalent events.  The weights are assigned according to how frequently they occur; numbers that occur more frequently, like three, are assigned a greater probability than more unusual numbers, like five. Using data from \citet{JimmyBoyd} that spans from 1980 to 2014, Table~\ref{tab:historical} shows historical college football data for selected point differentials.  It is possible that rule changes over time could warrant an approach that uses historical results from only certain years.  However, if the number of games is not too small, the trade-offs in which years to include will produce very similar results.  Further, a rule change to overtime added in 2021 will likely increase the probability that a game will finish with a point differential of two, but the increase should be relatively small.
\begin{table}[!ht]
\caption{Historical probabilities of selected point differentials.}
\begin{tabular}{r|c||r|c} 
Point & & Point & \\
Differential & Percentage & Differential & Percentage \\
\hline\hline
0 & 0\%   &  8 & 2.4\% \\
1 & 3.4\% &  9 & 1.2\% \\
2 & 2.7\% & 10 & 4.3\% \\
3 & 9.6\% & 11 & 2.3\% \\
4 & 3.9\% & 12 & 1.8\% \\
5 & 2.6\% & 13 & 1.8\% \\
6 & 2.9\% & 14 & 4.3\% \\
7 & 7.3\% & 15 & 1.1\% \\
\end{tabular}
\label{tab:historical}
\end{table}

An approach centered on historical data of all games works fairly well for National Football League (NFL) games mainly because the point spreads across all games do not have a large range. For example, the likelihood that a team favored to win by seven points actually wins by three is still relevant to that same question for a team favored by six points. According to {\tt sportsbettingdime.com}~\citep{spread}, there have only been nine NFL games since 1976 with point spreads larger than 20.  While it could be argued that information about how 20-point favorites perform is not relevant to how teams favored by 3 points will perform, this problem is exacerbated in college football, where spreads as large as 30 or 40 points are not uncommon.  In fact, in the 2021 college football season, more than 17\% (176) of the games had spreads larger than 20. Thus, rather than taking an aggregated approach, games that are similar to the game at hand provide the bettor with more relevant information for decision-making.

One solution to addressing the large range of point spreads in college football is using historical data to create conditional probabilities. Recall that for a standard $-110$ wager, the break-even point for bets is 52.4\%.  In the 2021 season, teams favored by between two and four points (inclusive) won games by more than one point 53.1\% of the time.  Thus, if a bettor believes that a team should be favored by three points, but the team is favored by only 1.5 points, a 53.1\% win expectancy translates to a 0.7\% edge over the house.  Classifying games into bins creates some challenges, including determining (1) the break-points to create the classifications or ``bins'' and (2) whether bins should overlap or be smoothed to account for a limited number of games in each bin.

\section{A New Approach for College Football Betting Edges}
\label{sec:new}
To address these challenges, a new method is proposed that is a hybrid of the methods in Sections~\ref{sec:basics} and~\ref{sec:historical}.  This new technique addresses the inclusion of non-relevant games in the aggregated historical data by using the data to optimally weight a normal distribution, which can then be centered at the bettor's point spread. The use of this new distribution alleviates the problems discussed when only binning the historical data.

The cumulative normal distribution with zero mean was applied to the historical data to find a probability distribution that best fits the historical probabilities.  Values half a point above and below a specific number are combined to represent that score differential.  Using a standard deviation of 21, the actual standard deviation of the games in the 2021 season, as discussed in Section~\ref{sec:normal}, fits the historical data well. However, a standard deviation of 22, which allows for extra variability, resulted in a closer match of the estimated probabilities to the historical data. Incremental changes to the standard deviation beyond 21 or 22 resulted in minimal changes to the final probabilities.
\begin{table}[H]
%\begin{tabular}{r|c c} 
\begin{tabular}{r|cc||r|cc}
%Point & Historical & Normal \\
%Differential & Probability & Probability \\
%\hline\hline
Point & Historical & Normal & Point & Historical & Normal \\
Differential & Probability & Probability & Differential & Probability & Probability \\ \hline\hline
%\hline\hline
%0  &  0\%  & 3.6\% \\
%1  & 3.4\% & 3.6\% \\
%2  & 2.7\% & 3.6\% \\
%3  & 9.6\% & 3.6\% \\
%4  & 3.9\% & 3.6\% \\
%5  & 2.6\% & 3.5\% \\
%6  & 2.9\% & 3.5\% \\
%7  & 7.3\% & 3.4\% \\
%8  & 2.4\% & 3.4\% \\
%9  & 1.2\% & 3.3\% \\
%10 & 4.3\% & 3.3\% \\
%11 & 2.3\% & 3.2\% \\
0  &  0\%  & 3.6\% &  6 & 2.9\% & 3.5\% \\
1  & 3.4\% & 3.6\% &  7 & 7.3\% & 3.4\% \\
2  & 2.7\% & 3.6\% &  8 & 2.4\% & 3.4\% \\
3  & 9.6\% & 3.6\% &  9 & 1.2\% & 3.3\% \\
4  & 3.9\% & 3.6\% & 10 & 4.3\% & 3.3\% \\
5  & 2.6\% & 3.5\% & 11 & 2.3\% & 3.2\% \\
\end{tabular}
\caption{Historical and fitted probabilities for selected point differentials in college football.}
\label{tab:historical2}
\end{table}

Next, for each point differential, a constant is found that scales the respective probability, translating it from the normal bell curve to a distribution where outcomes like three and seven have a larger probability, as the historical data suggest. The reasoning behind this step is that while a large favorite isn't likely to win by three, they are still more likely to win by three than by five.  The constant is found by dividing the normal probabilities in the third column of Table~\ref{tab:historical2} by the corresponding historical probabilities in column two.  For example, the area under a normal curve between 2.5 and 3.5 is 0.036.  However, the historical data suggests it should be 0.096.  Thus, the probability of a team winning by three when derived from a normal distribution should be multiplied by 2.7. This provides a framework to use the first method presented in Section~\ref{sec:normal}, but modifies that method, using the adjusted probabilities using historical information to take into account which football outcomes are more likely than others. Table~\ref{tab:mult} displays some of the multipliers so the reader can understand each number's relative importance compared to the unadjusted normal probability.
\begin{table}[H]
\begin{tabular}{r|c||r|c} 
Point         & & Point & \\ 
Differential  & Multiplier & 
Differential  & Multiplier \\
\hline\hline
0 &   0 &  6 & 0.8 \\
1 & 0.9 &  7 & 2.1 \\
2 & 0.7 &  8 & 0.7 \\
3 & 2.7 &  9 & 0.4 \\
4 & 1.1 & 10 & 1.3 \\
5 & 0.7 & 11 & 0.7 \\
\end{tabular}
\caption{Multipliers for selected point differentials.}
\label{tab:mult}
\end{table}

Next, a matrix is constructed for all differentials of 60 points or less.  Each row represents a game's score 
differential from $-60, -59, \ldots, -1, 0, 1, \ldots 59, 60$, where a negative value means the home team wins, and a positive value means the away team wins.  Each column represents the bettor's point spread from $-39$ to 39.
If $s$ is the score differential in a particular row, then a cell is the probability of being in the interval $(s - 0.5, s + 0.5)$, where the mean of the normal distribution is the bettor's point spread, and the standard deviation is 15.  Each cell is then multiplied by the appropriate weight for that score differential; see Table~\ref{tab:mult} for multipliers of select point differentials.  The final step for each column is to sum the numbers in that column and then divide each cell by the sum.  The result is that each column is now conditional probability distribution.  Specifically, each cell represents the conditional probability that the home team wins by exactly that score differential, given the bettor's point spread.

Because most systems project non-whole number results, the final cover probability can be computed via interpolation. For example, if a system indicates a team will win by 2.3, the method takes 30\% of the cover probability using the team projected as a two-point favorite and 70\% of the cover probability using the team projected as a three-point favorite.  For the Baylor/Texas game introduced in Section~\ref{sec:normal}, interpolation returns the probability of Baylor covering the 2.5-point spread to be 53.2\%, illustrating the importance of key numbers in wagering on football.  In other words, the new method accounts for the reality that Baylor is more likely to win by three than by two. Because this new methodology says there's an 0.8\% edge betting Baylor $-2.5$, this could be considered a worthwhile investment. Baylor covered the spread in a 7-point victory.

Finally, for projected spreads above 40 points, there is very little historical data.  Consequently, there is relatively little confidence in the resulting probabilities.  Projected spreads of greater than 40 are fairly rare, though the exact frequency depends on the bettor's projection system.  Games in this realm tend to be avoided by most sports bettors unless there is some qualitative information available as to how such a blowout might be handled by each coaching staff.

\section{Online App for Computing the Edge}
\begin{figure}[h]
\centering
\includegraphics[width=0.65\textwidth]{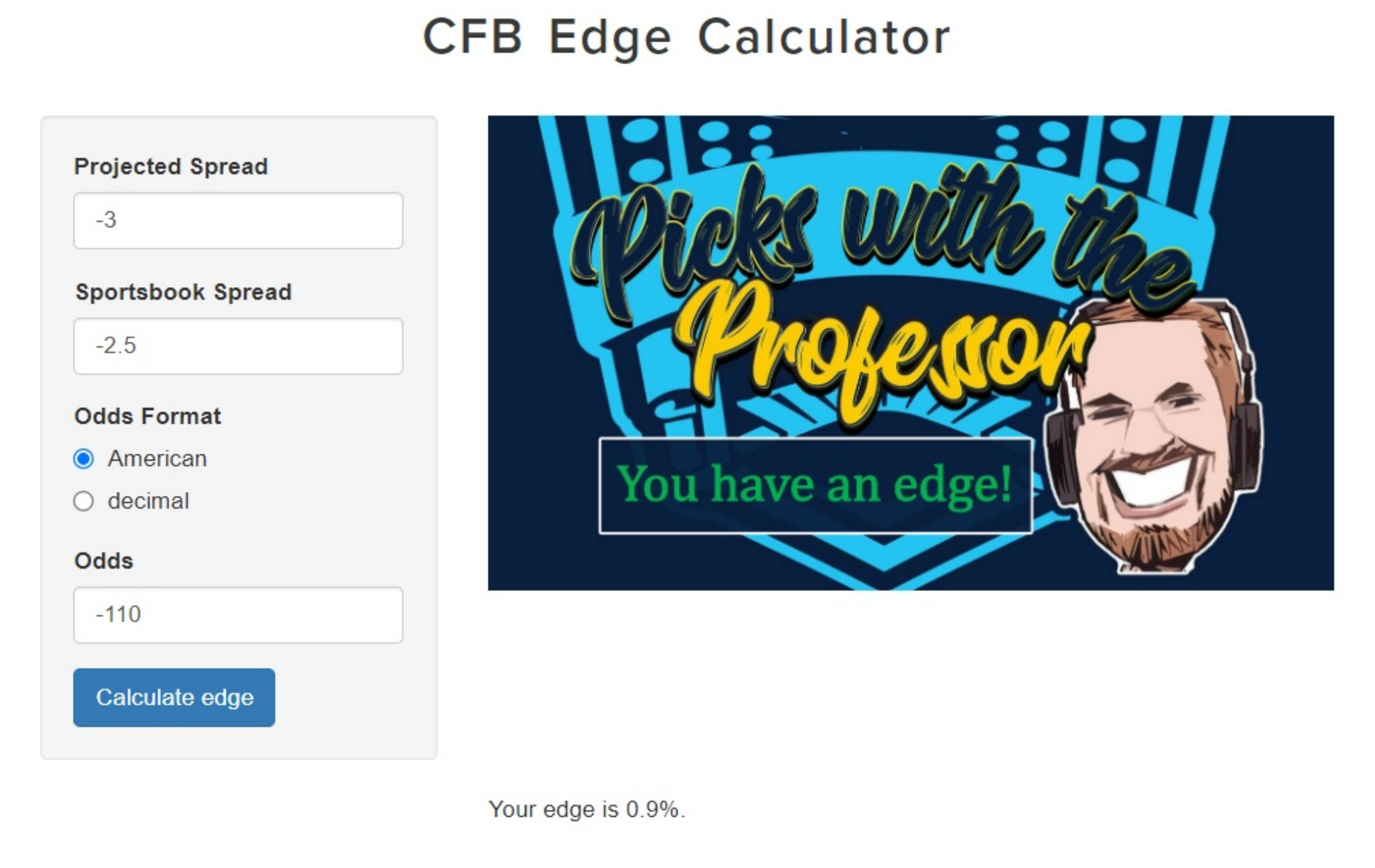}
\caption{Screenshot of ``Picks with the Professor'' online app for computing the edge of a specific bet.}
\label{fig:picks-default}
\end{figure}

An online tool that uses these probabilities to compute the edge is available online at {\tt www.pickswiththeprofessor.com/edge/cfb}.  A screenshot of the app is seen in Figure~\ref{fig:picks-default}. The user supplies four inputs, where all references are with respect to the team the bettor is interested in wagering on.

\begin{enumerate}
\item In the text box beneath {\tt Projected Spread}, the bettor's projected point spread is supplied, which can be determined from any source, like a complex mathematical model, or simply the bettor's belief.  The default value is $-3$.
\item In the text box beneath {\tt Sportsbook Spread}, the user supplies the betting point spread as provided by some sportsbook.
\item The third text box is where the {\tt Odds Format} is selected.  The default selection is {\tt American} odds, discussed in Section~\ref{sec:basics}.
\item Under {\tt Odds}, the user supplies the odds attached to the wager from Step 2.  The default is $-110$.
\end{enumerate}
Under the default scenario, the bettor has a 0.9\% edge.
It should be noted that there are some results that might seem counterintuitive. Most of them align with perception, but because the multiplier is so high for both three and seven, and to some extent, 10, a team that is believed to, on average, lose by eight actually has a 1.2\% edge if being bet at $+7.5$. However, for each bettor's projected point spread, it is confirmed mathematically that the expected value of each conditional distribution is very close to the projection, usually within one- or two-tenths of a point. This phenomenon occurs because of both a lack of symmetry and a lack of smoothness when imposing the multipliers to increase or decrease common and uncommon final score outcomes.

\section{Discussion}
\label{sec:future}
Some sports bettors place bets based on their instinct. However, using some mathematical modeling, perhaps in combination with other research, improves betting outcomes.  While transforming the results of a model into whether a wager is warranted is easy in moneyline sports, point spreads add a complication.  Point spreads force the bettor to decide the threshold for a profitable wager in a manner that is not straightforward.  The work here details a new approach to converting point differentials between a projected point spread and the actual spread into probabilities that can be very useful to a sports bettor.  A publicly available online tool on the ``Picks with the Professor'' website provides the percent edge a bettor has by using the spread and odds available along with the projected spread that the bettor believes to be the truth.

Further research is needed to understand how these probabilities perform with actual game data. Further, it is believed that probabilities over a certain threshold could produce misleading results because the projection model may not be able to account for every facet of an upcoming game, e.g., key player injuries.  While the 2021 college football season provides some insight into these issues, the lingering effects of the COVID-19 pandemic and the unique aspect of sixth-year seniors make the results less than ideal as a benchmark.  Data from the 2022 season should provide more clarity and a better evaluation of the projection systems.
\newpage
\bibliographystyle{apalike}
\bibliography{refs}
\end{document}